\documentclass[12pt, draftclsnofoot, onecolumn]{IEEEtran}

\usepackage{amssymb}
\usepackage{amsmath}
\usepackage{cite}
\usepackage{url}
\usepackage{empheq}
\usepackage{xcolor}
\usepackage{graphicx}
\usepackage{subfigure}
\usepackage{enumitem}
\usepackage{fancyhdr}
\usepackage{mdwmath}	
\usepackage{mdwtab}
\usepackage{caption}
\usepackage{amsthm}
\usepackage{algorithm}
\usepackage{algorithmic}

\newtheorem{lemma}{Lemma}

\newtheorem{theorem}{Theorem}
\newtheorem{corollary}{Corollary}
\newtheorem{assumption}{Assumption}
\newtheorem{definition}{Definition}

\newcommand{\eqr}[1]{(\ref{#1})}
\newcommand{\fref}[1]{Fig.~\ref{#1}}

\begin{document}
\title{User Clustering for Coexistence between Near-field and Far-field Communications}
\author{Kaidi~Wang,~\IEEEmembership{Member,~IEEE,}
Zhiguo~Ding,~\IEEEmembership{Fellow, IEEE,}
and George~K.~Karagiannidis,~\IEEEmembership{Fellow, IEEE}
\thanks{K. Wang is with the Department of Electrical and Electronic Engineering, The University of Manchester, Manchester, UK (email: kaidi.wang@ieee.org).}
\thanks{Z. Ding is with the Department of Computer and Information Engineering, Khalifa University, Abu Dhabi, UAE, and the Department of Electronic and Electrical Engineering, The University of Manchester, Manchester, UK (email: zhiguo.ding@ieee.org).}
\thanks{G. K. Karagiannidis is with Department of Electrical and Computer Engineering, Aristotle University of Thessaloniki, Greece and also with Artificial Intelligence \& Cyber Systems Research Center, Lebanese American University (LAU), Lebanon (email: geokarag@auth.gr).}}
\maketitle
%%%%%%%%%%%%%%%%%%%%%%%%%%%%%%%%%%%%%%%%%%%%%%%%%
%%%%%%%%%%%%%%%%%%%%%%%%%%%%%%%%%%%%%%%%%%%%%%%%%
\setlength{\abovedisplayskip}{4pt}
\setlength{\belowdisplayskip}{4pt}
\begin{abstract}
This letter investigates the coexistence between near-field (NF) and far-field (FF) communications, where multiple FF users are clustered to be served on the beams of legacy NF users, via non-orthogonal multiple access (NOMA). Three different successive interference cancellation (SIC) decoding strategies are proposed and a sum rate maximization problem is formulated to optimize the beam assignment and decoding order. The beam assignment problem is further reformulated as an overlapping coalitional game, which facilitates the design of the proposed clustering algorithm. The optimal decoding order in each cluster is also derived, which can be integrated into the proposed clustering. Simulation results demonstrate that the proposed clustering algorithm is able to significantly improve the sum rate of the considered system, and the developed strategies achieve different trade-offs between sum rate and fairness.
\end{abstract}
%%%%%%%%%%%%%%%%%%%%%%%%%%%%%%%%%%%%%%%%%%%%%%%%%
%%%%%%%%%%%%%%%%%%%%%%%%%%%%%%%%%%%%%%%%%%%%%%%%%
\section{Introduction}
In the sixth generation (6G) wireless networks, as the number of antenna elements increases, Rayleigh distance is significantly increased, which motivates the coexistence between near-field (NF) and far-field (FF) communication \cite{liu2023near, zhang2023nf}. Unlike conventional FF communications, which are based on beam steering channel models, NF communications rely on spherical-wave channel models, which enable beam focusing \cite{zhang2022nf}. As pointed out in \cite{zhang2022nf} and \cite{ding2023nf1}, the beams preconfigured for multiple NF users can be exploited to serve additional FF users in the same angular direction based on non-orthogonal multiple access (NOMA) schemes, where power allocation can be leveraged to suppress multi-access interference. Furthermore, it was shown in \cite{ding2023nf2} that the imperfect beamforming design for NF users can also provide opportunities for the implementation of NOMA. However, for the NOMA assisted coexistence between NF and FF communications, each FF user needs to utilize the appropriate beams for transmission and regard the signal transmitted through other beams as interference. Therefore, it is necessary to develop efficient user clustering algorithms and flexible continuous interference cancellation (SIC) decoding strategies. Although NOMA based NF communications have been studied in existing works \cite{ding2023nf1, zuo2023non}, this problem has not been well addressed. Specifically, the authors of \cite{ding2023nf1} studied power allocation with fixed user clustering, where a simple SIC decoding strategy was adopted at FF users, treating all other signals as interference. In \cite{zuo2023non}, user pairing was employed to support two-user NOMA transmission according to the different quality-of-service (QoS) requirements.

In order to explore the coexistence between NF and FF communications, this work studies the utilization of the legacy NF space division multiple access (SDMA) networks. Specifically, multiple FF users are included to perform NOMA schemes based on the generated beams for NF users. By decoding and eliminating inter- and/or intra-beam interference, three different SIC decoding strategies are proposed to balance complexity and performance in practical scenarios. Then, a sum rate maximization problem for FF users is formulated under the condition of satisfying the QoS of NF users, in which beam assignment and decoding order design are jointly investigated. Due to the fact that it is beneficial for each FF user to utilize multiple beams, beam assignment is formulated as an overlapping coalitional game which facilitates the design of the proposed user clustering algorithm\footnote{It is worth noting that the user clustering studied in this work is more challenging since i) the utility of any FF user in a cluster is partially decided by the other clusters; and ii) the utility of any cluster is affected by the permutation of FF users within it.}. Moreover, in each cluster, the optimal SIC decoding order for the intra-beam interference is derived. The simulation results show that the proposed solution can outperform the considered benchmarks in terms of sum rates, and the proposed SIC decoding strategies can be applied to different scenarios and objectives.
%%%%%%%%%%%%%%%%%%%%%%%%%%%%%%%%%%%%%%%%%%%%%%%%%
%%%%%%%%%%%%%%%%%%%%%%%%%%%%%%%%%%%%%%%%%%%%%%%%%
\section{System Model and Problem Formulation}
%%%%%%%%%%%%%%%%%%%%%%%%%%%%%%%%%%%%%%%%%%%%%%%%%
\subsection{System Model}
Consider a downlink SDMA system with one base station (BS), $K$ NF users, and $N$ FF users, where the BS is equipped with a $L$-antenna uniform linear array, and each user is equipped with a single antenna\footnote{The NOMA based coexistence considered in this paper can be utilized to serve add-on FF users in legacy NF communication scenarios, reduce the system overhead of beamforming re-configuration in networks with mobile FF users, or support overloaded systems where the number of users exceeds the number of antennas.}. The number of NF users is assumed to be less than or equal to the number of antennas at the BS, i.e., $K\le L$. The collections of all NF and FF users are $\mathcal{K}=\{1,2,\cdots,K\}$ and $\mathcal{N}=\{1,2,\cdots, N\}$, respectively. 

The channel vector of NF user $k$ is defined based on the spherical-wave propagation model \cite{zhu2022electromagnetic, zhang2022near}, as follows:
\begin{equation}
\mathbf{h}_k = \alpha_k\left[e^{-j\frac{2\pi}{\lambda}|\boldsymbol{\psi}_k^{\mathrm{NF}}-\boldsymbol{\psi}_1|} \quad \cdots \quad e^{-j\frac{2\pi}{\lambda}|\boldsymbol{\psi}_k^{\mathrm{NF}}-\boldsymbol{\psi}_L|}\right]^T,
\end{equation}
where $\alpha_k$ is the free space path loss, $\lambda$ is the wavelength, and $\boldsymbol{\psi}_k^{\mathrm{NF}}$ and $\boldsymbol{\psi}_l, \forall l \in \{1, 2, \cdots, L\}$ are the locations of NF user $k$ and the $l$-th element of the array, respectively. Specifically, the free space path loss is defined by $\alpha_k=\frac{c}{4\pi f_c|\boldsymbol{\psi}_k^{\mathrm{NF}}-\boldsymbol{\psi}_0|}$, where $c$, $f_c$, and $\boldsymbol{\psi}_0$ denote the speed of light, the carrier frequency, and the coordinate of the array center, respectively. By utilizing the beam steering vector, the channel vector of FF user $n$ can be modeled below:
\begin{equation}
\mathbf{g}_n\!\!=\!\alpha_n e^{-\!j\!\frac{2\pi}{\lambda}\!|\boldsymbol{\psi}_n^{\mathrm{FF}}\!-\boldsymbol{\psi}_1\!|}\!\!\left[1 \!\!\quad\!\! e^{-\!j\!\frac{2\pi d}{\lambda}\!\sin\!\theta_n} \!\!\quad\!\! \cdots \!\!\quad\!\! e^{-\!j\!\frac{2\pi d}{\lambda}\!(L-\!1)\!\sin\!\theta_n}\!\right]^T,
\end{equation}
where $\boldsymbol{\psi}_n^{\mathrm{FF}}$ is the location of FF user $n$, and $\theta_n$ is the angle of departure. It is assumed that perfect channel state information of NF and FF users is available at the BS, and beamforming is achieved through advanced beam training strategies, such as the one shown in \cite{zhang2022bt}. At the BS, zero-forcing beamforming is adopted based on the channel vectors of NF users, as shown in follows:
\begin{equation}
[\mathbf{p}_1 \!\quad\! \!\cdots\! \!\quad\! \mathbf{p}_K]=\mathbf{H}(\mathbf{H}^H\mathbf{H})^{-1},
\end{equation}
where $\mathbf{p}_k$ is the normalized beamforming vector, i.e., $\mathbf{p}_k^H\mathbf{p}_k=1, \forall k$, and $\mathbf{H}=[\mathbf{h}_1 \!\!\quad\! \!\!\cdots\!\! \!\quad\!\! \mathbf{h}_K]$. In particular, $K$ beams are generated to serve the users, and the signals for FF users is transmitted utilizing these beams. In this case, the NF users can enjoy the inter-beam interference-free transmission, while each FF user can utilize multiple beams to increase its individual data rate. The signal transmitted from the BS is given by
\begin{equation}
\mathbf{x}=\sum_{k=1}^K \mathbf{p}_k\sqrt{p_k}s_k^{\mathrm{NF}}+\sum_{k=1}^K \mathbf{p}_k\sum_{n=1}^N \sqrt{p_{k,n}}s_n^{\mathrm{FF}},
\end{equation}
where $p_k$ is the transmit power of NF user $k$, $p_{k,n}$ is the transmit power of FF user $n$ on beam $k$, and $s_k^{\mathrm{NF}}$ and $s_n^{\mathrm{FF}}$ are the signals for NF user $k$ and FF user $n$, respectively. The transmit power should satisfy $p_k\!+\!\sum_{n=1}^Np_{k,n}=P_t$, where $P_t$ is the transmit power budget for each beam. The received signal at NF user $k$ is given by
\begin{equation}
y_k=\mathbf{h}_k^H\mathbf{p}_k\sqrt{p_k}s_k^{\mathrm{NF}}+\mathbf{h}_k^H\mathbf{p}_k\!\sum_{n=1}^N\sqrt{p_{k,n}}s_n^{\mathrm{FF}}+n_k,
\end{equation}
where $n_k$ is the additive noise. Since NF users are not affected by inter-beam interference, and are expected to have better channel conditions than FF users, SIC techniques can be employed. That is, in each beam, the NF user can attempt to decode and remove all FF user signals. Assuming that the signals of FF users are decoded from $N$ to $1$, the data rate of NF user $k$ to decode the signal of FF user $n$ is given by
\begin{equation}
R_{k,n}^{\mathrm{NF}-\mathrm{FF}}\!\!=\!\log_2\!\!\left(\!\!1\!\!+\!\!\frac{p_{k,n}|\mathbf{h}_k^H\mathbf{p}_k|^2}{p_k|\mathbf{h}_k^H\mathbf{p}_k|^2\!\!+\!\!\sum_{i=1}^{n-1}\!p_{k,i}|\mathbf{h}_k^H\mathbf{p}_k|^2\!\!+\!\sigma^2}\!\!\right),\!
\end{equation}
where $\sigma^2$ is the variance of the noise. If the signals of all FF users are removed, or there is no FF user using the beam of NF user $k$,  the achievable data rate of NF user $k$ is given by
\begin{equation}
R_k^{\mathrm{NF}}=\log_2\left(1+\frac{p_k|\mathbf{h}_k^H\mathbf{p}_k|^2}{\sigma^2}\right).
\end{equation}
%%%%%%%%%%%%%%%%%%%%%%%%%%%%%%%%%%%%%%%%%%%%%%%%%
\subsection{SIC Decoding Design}
At FF user $n$, all signals for NF and FF users can be received, as shown in follows:
\begin{equation}
y_n=\mathbf{g}_n^H\sum_{k=1}^K\mathbf{p}_k\sqrt{p_k}s_k^{\mathrm{NF}}\!+\mathbf{g}_n^H\sum_{k=1}^K\mathbf{p}_k\!\sum_{n=1}^N\!\sqrt{p_{k,n}}s_n^{\mathrm{FF}}\!+n_n.
\end{equation}
It shows that each FF user suffers from both inter- and intra-beam interference simultaneously. In this case, we propose the following three SIC decoding strategies to remove the co-channel interference.

\subsubsection{SIC Decoding Strategy 1}
In case each FF user utilizes more than one beam, it will receive multiple desired signals. According to \cite{ding2023nf1}, the FF user can decode its signals by treating the signals of all other users as the interference. Assuming FF users decode signals from beam $K$ to beam $1$, the data rate of FF user $n$ to decode its signal transmitted through beam $k$ is
\begin{align}
\tilde{R}_{k,n}^{\mathrm{FF}(1)}=\log_2\!\!\left[\!1\!\!+\!\!\frac{p_{k,n}g_{k,n}}{\sum_{k=1}^K\!(p_k\!\!+\!\!\sum_{i=1,i\neq n}^N\!p_{k,i})g_{k,n}\!\!+\!\!\sum_{i=1}^{k-1}\!p_{i,n}g_{i,n}\!\!+\!\sigma^2}\!\right],
\end{align}
where $g_{k,n}\triangleq |\mathbf{g}_n^H\mathbf{p}_k|^2$.

\subsubsection{SIC Decoding Strategy 2}
Similar to NF users, strong FF users with better channel conditions and/or weaker interference are also able to decode the signals for weak FF users. In this case, a decoding order should be designed for each beam. With a given decoding order $Q_k$, the signals of FF users are decoded from the last to the first. The data rate of FF user $n$ to decode the signal of FF user $m$ on beam $k$ is given by
\begin{align}
\hat{R}_{k,n\to m}^{\mathrm{FF}-\mathrm{FF}(2)}=\log_2\!\!\left[\!1\!+\!\frac{p_{k,m}g_{k,n}}{P_t\!\sum_{i=1, i\neq k}^K\!g_{i,n}\!\!+\!(p_k\!\!+\!\!\sum_{i=1}^{m-1}\!p_{k,i})g_{k,n}\!\!+\!\sigma^2}\!\right].
\end{align}
After partially removing the signals of FF users with the same beam, the signals of other users are treated as interference, and the data rate of FF user $n$ with beam $k$ can be expressed as
\begin{equation}
\hat{R}_{k,n}^{\mathrm{FF}(2)}\!\!=\!\log_2\!\!\left[\!1\!+\!\frac{p_{k,n}g_{k,n}}{P_t\!\sum_{i=1,i\neq k}^K\!g_{i,n}\!\!+\!(p_k\!\!+\!\!\sum_{i=1}^{n-1}\!p_{k,i})g_{k,n}\!\!+\!\sigma^2}\!\right],
\end{equation}
In order to guarantee the successful decoding, the data rate of FF user $n$ on beam $k$ with decoding order $Q_k$ is given by
\begin{equation}
\tilde{R}_{k,n}^{\mathrm{FF}(2)}\!=\!\min\!\left\{\!\hat{R}_{k,m\to n}^{\mathrm{FF}-\mathrm{FF}(2)}, \hat{R}_{k,n}^{\mathrm{FF}(2)}\big|\forall Q_k(m)\!<\!Q_k(n)\!\right\}.
\end{equation}

\subsubsection{SIC Decoding Strategy 3}
The aforementioned two decoding strategies can be combined to form SIC decoding strategy 3. That is, each FF user can decode and remove the signals of weak users on all beams following its beam. For FF user $n$, in order to decode the signal of FF user $m$ on beam $k$, it can remove the signals of all users $m'$ on all beams from $K$ to $k$,  where $m'>m>n$. The rate of FF user $n$ to decode the signal of FF user $m$ on beam $k$ is
\begin{align}
\hat{R}_{k,n\to m}^{\mathrm{FF}-\mathrm{FF}(3)}=\log_2\!\!\left[\!1\!+\!\frac{p_{k,m}g_{k,n}}{P_t\!\sum_{i=1}^{k-1}\!g_{i,n}\!+\!\sum_{i=k}^K(p_i\!+\!\!\sum_{j=1}^{m-1}\!p_{i,j}\!)g_{i,n}\!\!+\!\sigma^2}\!\right].
\end{align}
After that, FF user $n$ can decode its own signal on beam $k$ at the following rate:
\begin{equation}
\hat{R}_{k,n}^{\mathrm{FF}(3)}\!\!=\!\log_2\!\!\left[\!1\!\!+\!\!\frac{p_{k,n}g_{k,n}}{P_t\!\sum_{i=1}^{k-1}\!g_{i,n}\!\!+\!\!\sum_{i=k}^K\!(p_i\!+\!\!\sum_{j=1}^{n-1}\!p_{i,j})g_{i,n}\!\!+\!\sigma^2}\!\right].
\end{equation}
Similarly, the achievable data rate of FF user $n$ is limited by the stronger FF users, as shown in the following
\begin{equation}
\tilde{R}_{k,n}^{\mathrm{FF}} = \min\!\left\{\!\hat{R}_{k,m\to n}^{\mathrm{FF}-\mathrm{FF}(3)}, \hat{R}_{k,n}^{\mathrm{FF}(3)}\big|\forall Q_k(m)\!<\!Q_k(n)\!\right\}.
\end{equation}

With any SIC decoding strategy, the data rate of FF user $n$ utilizing beam $k$ should be limited by the corresponding NF user, as shown in follows:
\begin{equation}
R_{k,n}^{\mathrm{FF}(i)}=\min\left\{R_{k,n}^{\mathrm{NF}-\mathrm{FF}}, \tilde{R}_{k,n}^{\mathrm{FF}(i)}\right\}, \forall i\in \{1,2,3\}.
\end{equation}
For simplicity, the superscript $(i)$ denoting the SIC decoding strategy  is omitted in the remainder of this letter.
%%%%%%%%%%%%%%%%%%%%%%%%%%%%%%%%%%%%%%%%%%%%%%%%%
\subsection{Problem Formulation}
In this letter, a sum rate maximization problem is formulated for the FF users, under the condition of guaranteeing the QoS of the NF users. To indicate the utilization of beams, a binary beam assignment indicator $x_{k,n}$ is defined based on the transmit power, as shown below:
\begin{equation}
x_{k,n} = \left\{
\begin{array}{ll}1,&  \text{if}\quad p_{k,n} > 0,\\
0, & \text{if}\quad p_{k,n} = 0.\\
\end{array}\right.
\end{equation}
That is, $x_{k,n}=1$ indicates that FF user $n$ is using the beam of NF user $k$; otherwise $x_{k,n}=0$. Therefore, the problem can be presented as follows:
\begin{align}
\min_{\mathbf{X},\mathbf{Q}}\quad &\sum_{n=1}^N\sum_{k=1}^Kx_{k,n}R_{k,n}^{\mathrm{FF}} \label{problem}\\
\textrm{s.t.} \quad & R_k^{\mathrm{NF}}\ge R_{\mathrm{min}}, \forall k\in\mathcal{K}, \tag{\ref{problem}a}\\
& \sum\nolimits_{k=1}^K x_{k,n} \ge 1, \forall n\in\mathcal{N},  \tag{\ref{problem}b}\\
& x_{k,n} \in \{0, 1\}, \forall k\in\mathcal{K}, \forall n\in\mathcal{N}, \tag{\ref{problem}c}
\end{align}
where $\mathbf{X}$ and $\mathbf{Q}$ are the collections of all beam assignment indicators and SIC decoding orders, respectively. Note that constraint (\ref{problem}a) includes the target data rate of NF users, i.e., $R_{\mathrm{min}}$, and constraint (\ref{problem}b) implies that each FF user should utilize at least one beam.

%%%%%%%%%%%%%%%%%%%%%%%%%%%%%%%%%%%%%%%%%%%%%%%%%
%%%%%%%%%%%%%%%%%%%%%%%%%%%%%%%%%%%%%%%%%%%%%%%%%
\section{User Clustering and Decoding Order Design}
In this section, the proposed problem is solved by iteratively executing the proposed user clustering algorithm and the designed SIC decoding order.
%%%%%%%%%%%%%%%%%%%%%%%%%%%%%%%%%%%%%%%%%%%%%%%%%
\subsection{Coalitional Game based User Clustering}
The beam assignment problem formulated in \eqr{problem} can be viewed as an overlapping coalitional game $(\mathcal{N}, U, \mathcal{S})$, in which $N$ FF users form a structure $\mathcal{S}$ in order to improve the utility $U$. The structure is a collection of  $K$ overlapping clusters, i.e., $\mathcal{S}=\{S_1, S_2, \cdots, S_K\}$, where each cluster $S_k$ includes one NF user and multiple FF users.  Without loss of generality, the NF user in cluster $S_k$ is denoted by $k$. In the considered coalition game, the FF users can request to join or leave clusters based on the sum rate, while the corresponding NF user can accept or reject this request based on the QoS constraints. The coalition utility is defined as follows:
\begin{equation}
U(\mathcal{S})=\sum_{k\in\mathcal{K}}\sum_{n\in S_k} R_{k,n}^{\mathrm{FF}}.
\end{equation}

During the game, the requests of FF users are based on the merge-and-split strategy \cite{han2012game, chen2021game}, as shown in follows:
\begin{definition}
(\textbf{Merge Rule}) With a given cluster $S_k$ in $\mathcal{S}$, any FF user $n \notin S_k$ tends to merge if and only if $U(\mathcal{S})< U(\mathcal{S}\backslash S_k\cup S_k')$, where $S_k'=S_k\cup \{n\}$.
\end{definition}
\begin{definition}
(\textbf{Split Rule}) In structure $\mathcal{S}$, any FF user $n \in S_k$ tends to split if and only if $U(\mathcal{S})< U(\mathcal{S}\backslash S_k\cup S_k')$ and $\mathcal{S}\backslash S_k\cap\{n\}\neq\emptyset$, where $S_k'=S_k\backslash \{n\}$.
\end{definition}
Based on the merge and split rule, the FF user can make an application to join or leave a cluster, and the corresponding NF user must respond to the application. For the merge request, the NF user responds according to its individual data rate. Specifically, if the target data rate of NF user can be achieved after the FF user joins its cluster, the application can be accepted; otherwise, the application is rejected. For the merge request of FF user $n$, the strategy of NF user $k$ can be presented as follows:
\begin{equation}
\left\{
\begin{array}{ll}S_k \to S_k\cup \{n\},& \text{if}\!\quad\! R_k^\mathrm{NF}(S_k\cup\{n\}) \ge R_{\mathrm{min}},\\
S_k \to S_k , & \text{otherwise}.\\
\end{array}\right.
\end{equation}
Note that in the scenario considered, the transmit power of the NF user does not decrease when the FF user leaves the cluster, and hence, the data rate of the NF user is not reduced by this action. Therefore, the split requests are always approved.

Based on the strategies of FF and NF users, a user clustering algorithm is proposed in Algorithm \ref{alg1}. For Algorithm \ref{alg1}, the FF users start actions from $1$ to $N$.  If FF user $N$ completes the search action from $S_1$ to $S_K$, a new round of the main loop starts. In order to guarantee that the data rate of FF users is not reduced due to SIC decoding, once the structure $\mathcal{S}$ is changed, the designed SIC decoding order in Section III-B can be performed for all clusters.

\setlength{\textfloatsep}{0pt}
\begin{algorithm}[t]
\caption{Game Theory Based User Clustering Algorithm}
\label{alg1}
\begin{algorithmic}[1]
\STATE \textbf{Initialization:}
\STATE Initialize the structure of $\mathcal{S}$ by randomly dividing FF users into all clusters.
\STATE \textbf{Main Loop:}
\FOR{$n\in\mathcal{N}$}
\STATE FF user $n$ searches cluster $S_k\in\mathcal{S}$.
\IF{Merge Rule is satisfied}
\STATE FF user $n$ applies to join in cluster $S_k$.
\IF{$R_k^\mathrm{NF}(S_k\cup\{n\}) \ge R_{\mathrm{min}}$}
\STATE Set $\mathcal{S}= \mathcal{S}\backslash S_k\cup S_k'$, where $S_k'=S_k\cup \{n\}$.
\ENDIF
\ENDIF
\IF{Split Rule is satisfied}
\STATE FF user $n$ leaves cluster $S_k$.
\STATE Set $\mathcal{S}=\mathcal{S}\backslash S_k \cup S_k'$, where $S_k'=S_k\backslash\{n\}$.
\ENDIF
\ENDFOR
\STATE The main loop is repeated until no merge or split strategy can be performed in a complete round.
\end{algorithmic}
\end{algorithm}

\subsubsection{Complexity Analysis}
The computational complexity of the proposed user clustering algorithm is $\mathcal{O}(CKN)$, where $C$ is the number of the main loops. In each round, $N$ FF users have to search $K$ clusters, so $KN$ computations are performed. With a given number of rounds $C$, the computational complexity can be obtained.

\subsubsection{Convergence and Stability}
From any initial structure, the proposed algorithm is guaranteed to converge to a stable structure. Suppose that $\mathcal{S}_a$ and $\mathcal{S}_b$ are two adjacent structures, where $a<b$, from $\mathcal{S}_a$ to $\mathcal{S}_b $, there is a FF user who executes the merge or split strategy. Based on the definitions, the utility is strictly increasing i.e., $U(\mathcal{S}_a)<U(\mathcal{S}_b)$. With the finite number of FF users and clusters, the number of possible structures and the upper bound on the sum rate are also finite \cite{han2012game}. As a result, the structure can always converge to a final structure, which is also a stable structure.

%%%%%%%%%%%%%%%%%%%%%%%%%%%%%%%%%%%%%%%%%%%%%%%%%
\subsection{SIC Decoding Order Design}
 In order to improve the performance of SIC decoding strategies 2 and 3, the FF users should be ordered in each cluster. For any given cluster $S_k=\{k, 1,2,\cdots, |S_k|\}$, the SIC decoding order $Q_k$ is from $|S_k|$ to $1$. In other words, each FF user must decode and remove the signal of the FF user after it. Moreover, in the case that $m<n$, if FF user $m$ cannot decode the signal of another FF user $n$, the data rate of FF user $n$ should be reduced \cite{kaidi2019clustering}. As a result, the data rate of any FF user $n$ is limited by all FF users that decode its signals, as shown in the following equation:
\begin{equation}
\tilde{R}_{k,n}^{\mathrm{FF}}=\min\left\{\hat{R}_{k,m\to n}^{\mathrm{FF}-\mathrm{FF}} \big| \forall m \le n\right\}.
\end{equation}

When using SIC decoding strategy 2, the above equation can be transformed as follows:
\begin{align}\nonumber
\!\!\!\tilde{R}_{k,n}^{\mathrm{FF}(2)}&\!\!=\!\min\!\!\left\{\!\log_2\!\left(\!1\!+\!\frac{p_{k,n}}{A_{k,m}\!\!+\!p_k\!+\!\sum_{i=1}^{n-1}\!p_{k,i}}\!\right)\!\bigg| \forall m \le n\!\right\}\\
&=\!\log_2\!\!\left(\!\!1\!\!+\!\!\frac{p_{k,n}}{\max\{A_{k,m}^{(2)} |\forall m \!\le\! n\}\!+\!p_k\!\!+\!\!\sum_{i=1}^{n-1}\!p_{k,i}}\!\right),\!\!\!
\end{align}
where 
\begin{equation}
A_{k,m}^{(2)}\triangleq\frac{P_t\!\sum_{i=1, i\neq k}^K\!g_{i,m}\!\!+\!\sigma^2}{g_{k,m}}.
\end{equation}
In order to maximize the data rate of each FF user in cluster $S_k$, it should be guaranteed that $A_{k,n}^{(2)}$ is the maximum when decoding the signal of FF user $n$, i.e.,
\begin{equation}
\max\{A_{k,1}^{(2)}, A_{k,2}^{(2)}, \cdots, A_{k,n}^{(2)}\} = A_{k,n}^{(2)}, \forall n \in S_k.
\end{equation}
In this case, the data rate of FF user $n$ in cluster $S_k$ is not reduced due to the SIC decoding. As a result, the FF users in cluster $S_k$ should be ordered as follows:
\begin{align}\nonumber
&\max\left\{A_{k,m}^{(2)}\big|\forall m \le 1 \right\} \le \cdots \le \max\left\{A_{k,m}^{(2)}\big|\forall m \le |S_k|\right\}\\
 \Rightarrow &\frac{P_t\!\sum_{i=1, i\neq k}^K\!g_{i,1}\!\!+\!\sigma^2}{g_{k,1}} \!\le\!\cdots\!\le\!\frac{P_t\!\sum_{i=1, i\neq k}^K\!g_{i,|S_k|}\!\!+\!\sigma^2}{g_{k,|S_k|}}.
\label{order}
\end{align}
The above inequality indicates that in any cluster $S_k$, the FF users should be ordered based on inter-beam interference and channel gains. In particular, any FF user with the strong channel condition and/or the weak inter-beam interference is able to decode the signals of other FF users without reducing their data rates. As a result, the sum rate can be increased. Moreover, the data rates of NF user $k$ for decoding the signal of FF user $n$ is given by
\begin{equation}
R_{k,n}^{\mathrm{NF}-\mathrm{FF}}=\log_2\!\!\left(\!1\!+\!\frac{p_{k,n}}{A_k+p_k\!+\!\sum_{i=1}^{n-1}\!p_{k,i}}\right),
\end{equation}
where $A_k\triangleq\sigma^2/|\mathbf{h}_k^H\mathbf{p}_k|^2$. It is noted that allowing the NF user to decode  the signals of all FF users is consistent with the proposed decoding order, since there is no inter-beam interference at NF users.

With SIC decoding strategy 3, the data rate of any FF user $n$ in cluster $S_k$ can be rewritten as follows:
\begin{equation}
\tilde{R}_{k,n}^{\mathrm{FF}(3)}=\log_2\!\!\left(\!1\!+\!\frac{p_{k,n}}{\max\{A_{k,m\to n}^{(3)} | \forall m \le n\}}\!\right),
\end{equation}
where
\begin{equation}
A_{k,m\to n}^{(3)}\!\!=\!\frac{P_t\!\sum_{i=1}^{k-1}\!g_{i,m}\!\!+\!\!\sum_{i=k}^K(p_i\!+\!\!\sum_{j=1}^{n-1}\!p_{i,j}\!)g_{i,m}\!\!+\!\sigma^2}{g_{k,m}}.\!\!\!
\end{equation}
From the above equations, the optimal decoding order in SIC strategy 3 can be obtained in a similar way. However, the optimal decoding order in SIC strategy 3 is affected by the transmit power of other users, which makes the optimization of power allocation challenging. Therefore, in this letter, only the decoding order in \eqr{order} is adopted, which can be considered as a suboptimal solution to SIC strategy 3.
%%%%%%%%%%%%%%%%%%%%%%%%%%%%%%%%%%%%%%%%%%%%%%%%%
%%%%%%%%%%%%%%%%%%%%%%%%%%%%%%%%%%%%%%%%%%%%%%%%%
\section{Simulation Results}
In the simulations, a half-disc is considered, where the coordinate of BS is $\psi_0=(0,0)$. The NF users are randomly distributed in a half-ring with inner diameter $5$~m and outer diameter $21.2625$~m, and the FF users are randomly distributed in a half-ring with inner diameter $86.4054$~m and outer diameter $96.4054$~m. $L=64$, $f_c=28$~GHz,  $R_\mathrm{min}=0.2$ bits per channel use, and $\sigma^2=-80$~dBm. Moreover, random user clustering (UC) and random decoding order (DO) are included as the benchmarks.

\begin{figure}[t]
\centering{
\subfigure[Sum Rate]{\includegraphics[width=80mm]{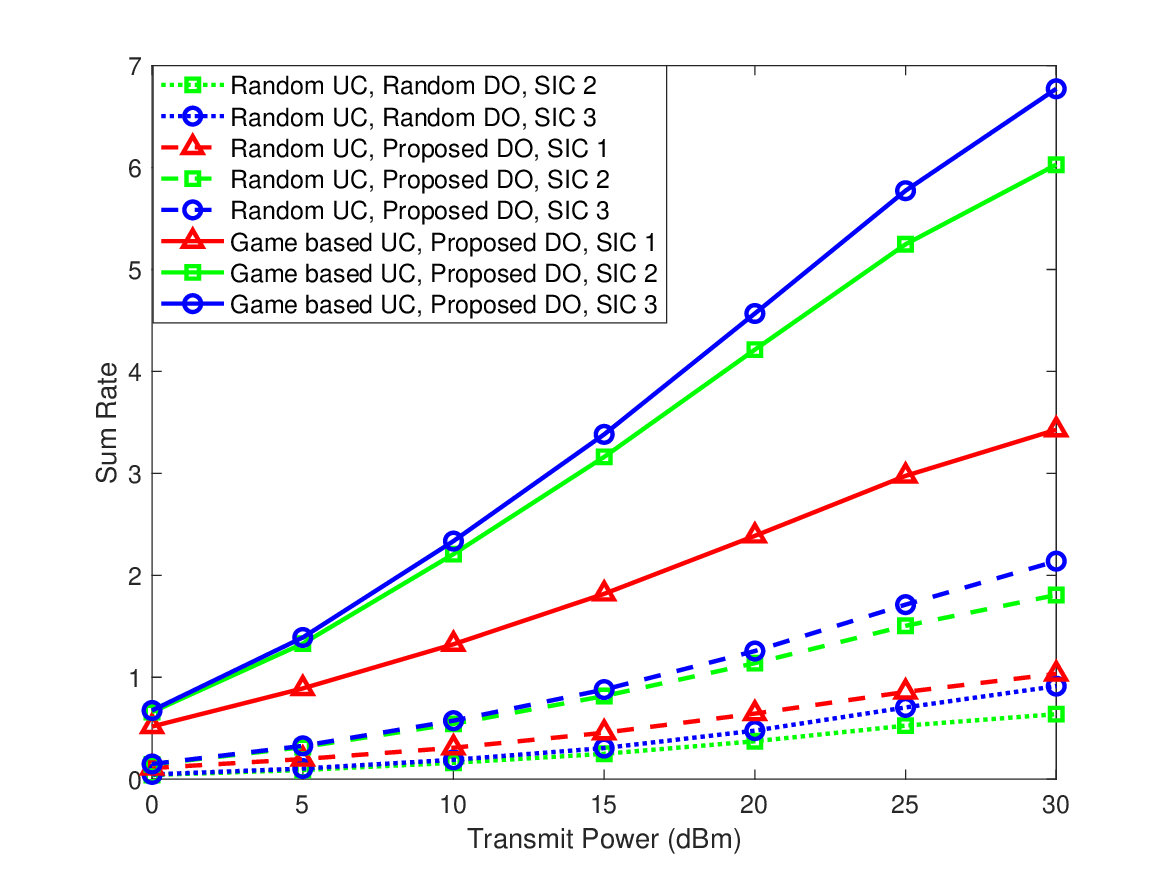}}
\subfigure[Jain's Fairness]{\includegraphics[width=80mm]{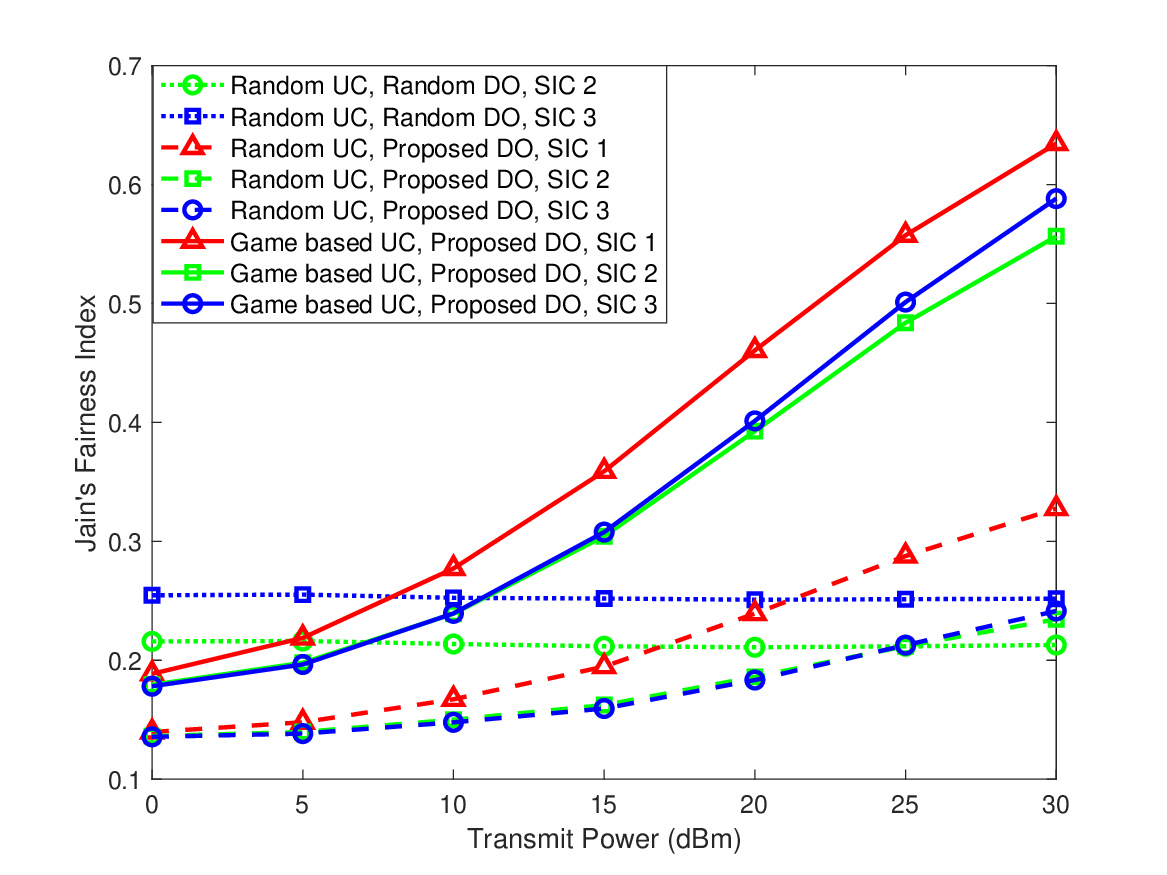}}}
\caption{The impact of transmit power. $L=64$, $K=5$, and $N=20$.}
\label{result1}
\end{figure}

\fref{result1} shows the performance of the considered system, where the Jain's fairness index is introduced and calculated by $(R_\mathrm{sum}^\mathrm{FF})^2/(NR_\mathrm{sum}^\mathrm{FF})$ , where $R_\mathrm{sum}^\mathrm{FF}=\sum_{k=1}^K\sum_{n=1}^NR_{k,n}^{\mathrm{FF}}$. It is shown that the proposed solution, including game based UC and DO, can significantly improve the sum rate, and the improvement increases with transmit power. Due to the fact that strong FF users can remove more signals with SIC strategy 2 and SIC strategy 3, these strategies can achieve significantly higher sum rates compared to the one studied in \cite{ding2023nf1}, i.e., SIC strategy 1. Moreover, it is worth pointing out that in SIC strategy 2 and SIC strategy 3, since the number of removed signals at FF users is different according to different channel gains, the fairness of these strategies is lower.

\begin{figure}[t]
\centering{\includegraphics[width=80mm]{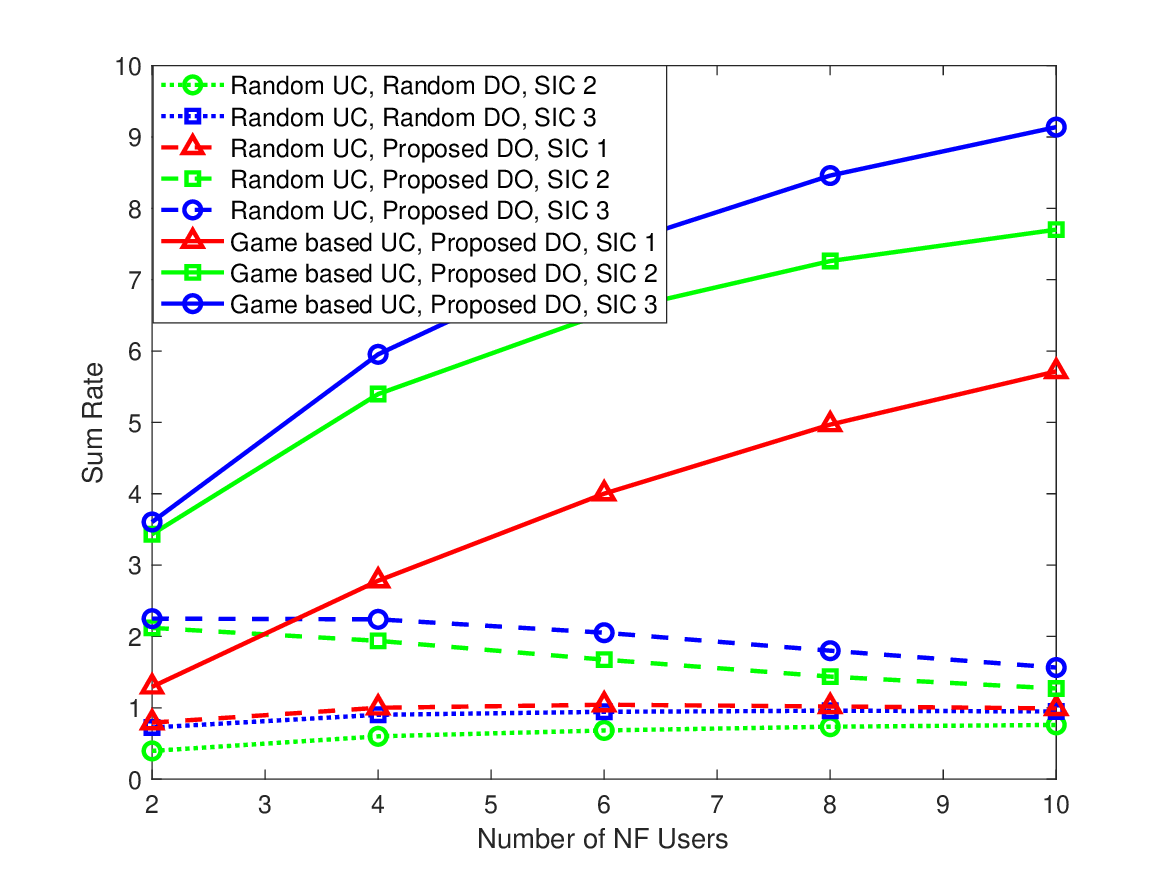}}
\caption{The impact of the number of NF users. $L=64$, $N=20$,  and $P_t=30$~dBm.}
\label{result2}
\end{figure}

\begin{figure}[t]
\centering{\includegraphics[width=80mm]{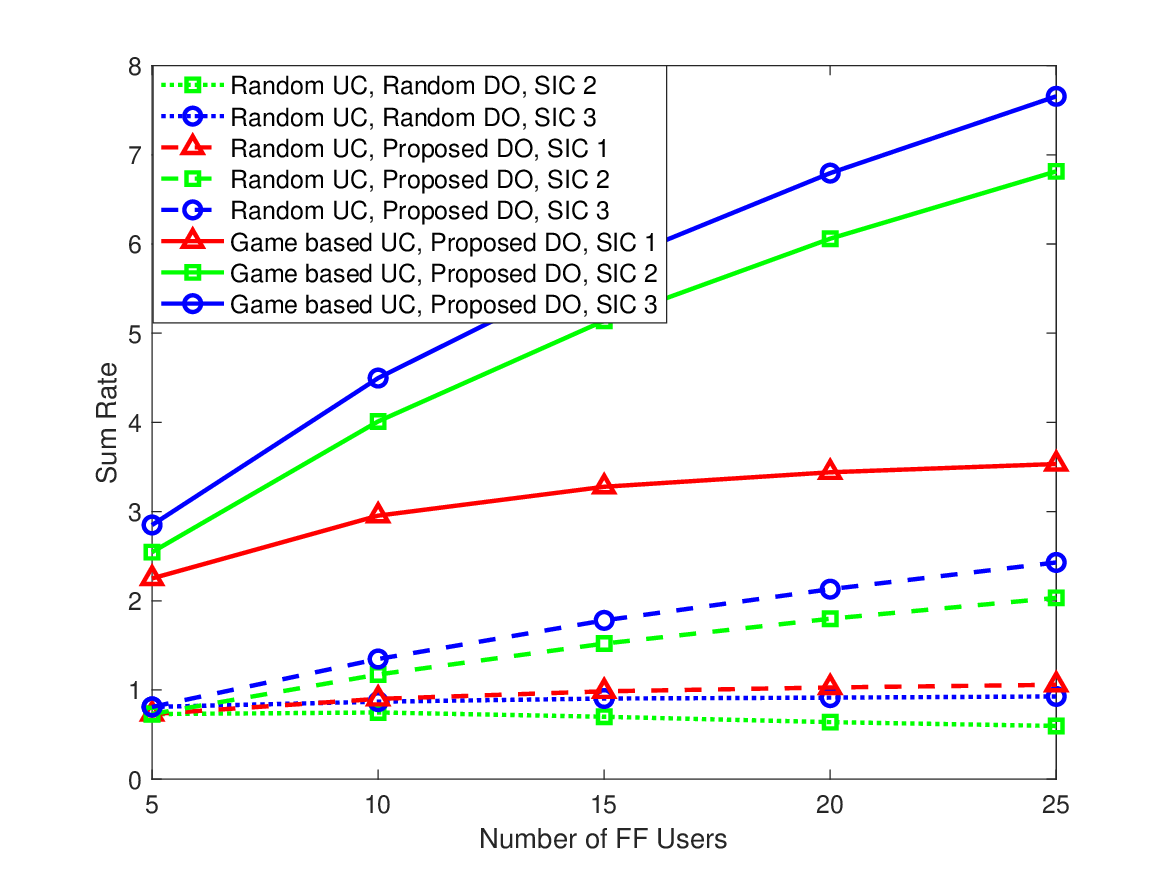}}
\caption{The impact of the number of FF users. $L=64$, $K=5$, and $P_t=30$~dBm.}
\label{result3}
\end{figure}

\fref{result2} and \fref{result3} demonstrate the sum rate with different numbers of NF users and FF users, respectively. Overall, the proposed solution allows the sum rate to increase with the number of users. When the number of NF users increases, the number of clusters is increased, and then the gap between SIC strategy 1 and SIC strategy 2 does not change. As the number of FF users increases, the number of FF users in each cluster increases, and the advantage of SIC strategies for decoding inter-beam interference becomes obvious. As a result, in \fref{result3}, the sum rate achieved by SIC strategies 2 and 3 is significantly increased. It is worth noting that SIC strategy 3 can always achieve the best performance, because it is a combination of the other two strategies.

\begin{figure}[t]
\centering{
\subfigure[Sum Rate]{\includegraphics[width=80mm]{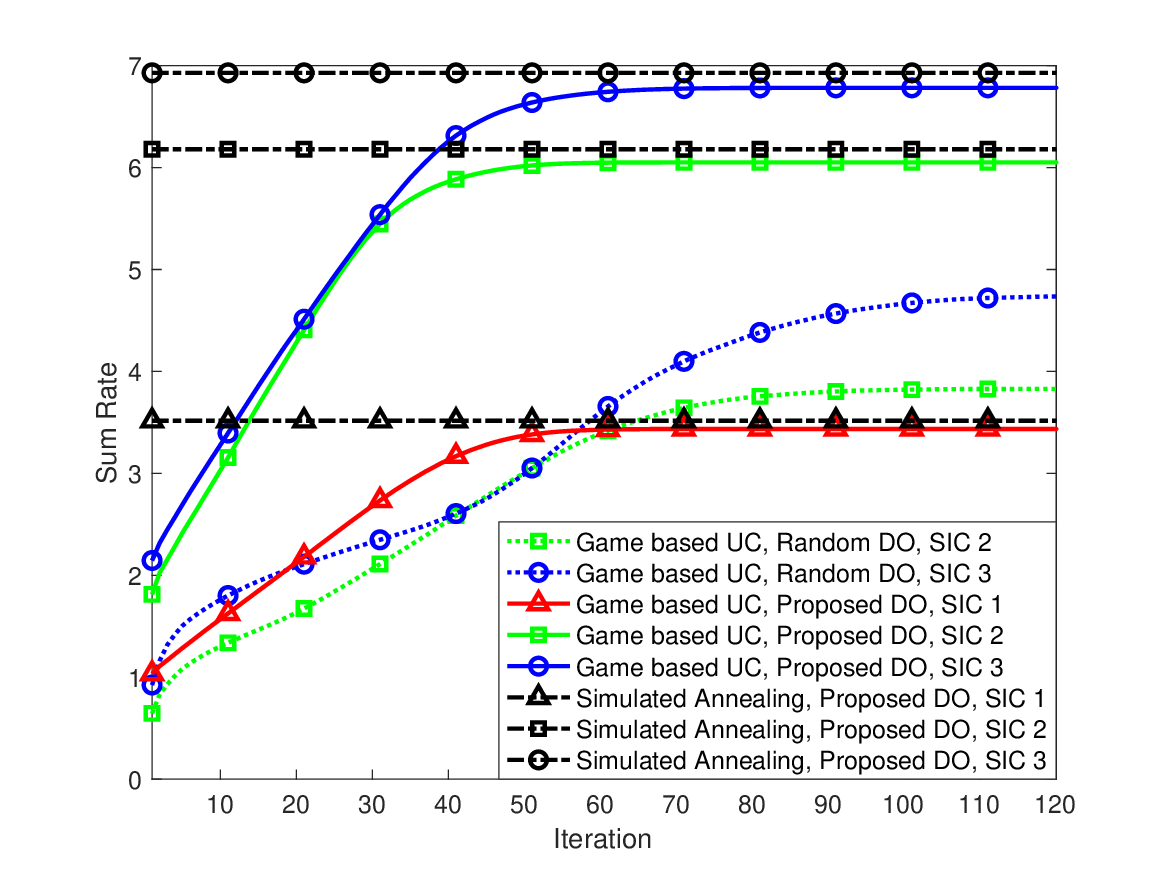}}
\subfigure[Average Rate of NF Users]{\includegraphics[width=80mm]{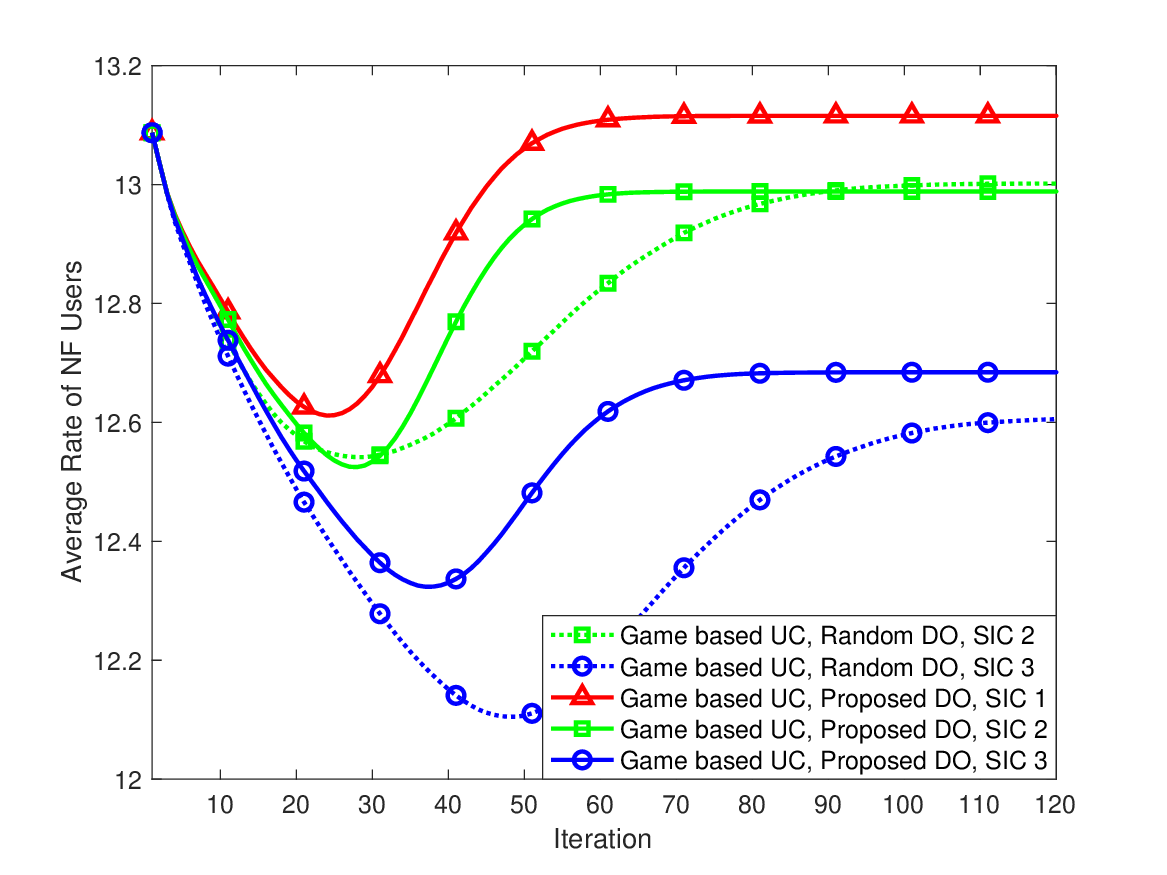}}}
\caption{The convergence of the user clustering algorithm. $L=64$, $K=5$, $N=20$, and $P_t=30$~dBm.}
\label{result4}
\end{figure}

In \fref{result4}, the analysis of convergence and stability is conducted, where simulated annealing is included as a benchmark. Note that although simulated annealing based algorithms can converge to a global optimal solution with sufficient iterations, its optimality is limited in the overlapping coalitional game. \fref{result4}(a) shows that the proposed user clustering algorithm can achieve approximately $98\%$ performance of the simulated annealing based algorithm with much lower complexity. It is also shown that the proposed algorithm needs more iterations to converge to a stable structure if random DO is used. In \fref{result4}(b), the change in the average rate of NF users shows that the improvement of the sum rate in the early stage is caused by the merging strategy, and then the splitting strategy plays a dominant role. Furthermore, compared with random DO, the proposed DO is more efficient, as it can achieve higher sum rates for FF users while ensuring that the average rates for NF users are at a similar or higher level.

%%%%%%%%%%%%%%%%%%%%%%%%%%%%%%%%%%%%%%%%%%%%%%%%%
%%%%%%%%%%%%%%%%%%%%%%%%%%%%%%%%%%%%%%%%%%%%%%%%%
\section{Conclusion}
In this letter, beam assignment and decoding order design were studied in a NOMA assisted NF communication system in order to maximize the sum rate of all FF users. A coalitional game based user clustering algorithm was developed, in which the designed decoding order can be executed to implement the proposed SIC decoding strategies. The improvement caused by the proposed solution and the characteristics of the provided SIC decoding strategies were shown in simulation results.
%%%%%%%%%%%%%%%%%%%%%%%%%%%%%%%%%%%%%%%%%%%%%%%%%
%%%%%%%%%%%%%%%%%%%%%%%%%%%%%%%%%%%%%%%%%%%%%%%%%
\bibliographystyle{IEEEtran}
\bibliography{KaidisBib}

\begin{thebibliography}{10}
\providecommand{\url}[1]{#1}
\csname url@samestyle\endcsname
\providecommand{\newblock}{\relax}
\providecommand{\bibinfo}[2]{#2}
\providecommand{\BIBentrySTDinterwordspacing}{\spaceskip=0pt\relax}
\providecommand{\BIBentryALTinterwordstretchfactor}{4}
\providecommand{\BIBentryALTinterwordspacing}{\spaceskip=\fontdimen2\font plus
\BIBentryALTinterwordstretchfactor\fontdimen3\font minus \fontdimen4\font\relax}
\providecommand{\BIBforeignlanguage}[2]{{%
\expandafter\ifx\csname l@#1\endcsname\relax
\typeout{** WARNING: IEEEtran.bst: No hyphenation pattern has been}%
\typeout{** loaded for the language `#1'. Using the pattern for}%
\typeout{** the default language instead.}%
\else
\language=\csname l@#1\endcsname
\fi
#2}}
\providecommand{\BIBdecl}{\relax}
\BIBdecl

\bibitem{liu2023near}
Y.~Liu, J.~Xu, Z.~Wang, X.~Mu, and L.~Hanzo, ``Near-field communications: What will be different?'' \emph{arXiv preprint arXiv:2303.04003}, 2023.

\bibitem{zhang2023nf}
Y.~Zhang, C.~You, L.~Chen, and B.~Zheng, ``Mixed near- and far-field communications for extremely large-scale array: An interference perspective,'' \emph{{IEEE} Commun. Lett.}, vol.~27, no.~9, pp. 2496--2500, 2023.

\bibitem{zhang2022nf}
H.~Zhang, N.~Shlezinger, F.~Guidi, D.~Dardari, M.~F. Imani, and Y.~C. Eldar, ``Beam focusing for near-field multiuser {MIMO} communications,'' \emph{{IEEE} Trans. Wireless Commun.}, vol.~21, no.~9, pp. 7476--7490, 2022.

\bibitem{ding2023nf1}
Z.~Ding, R.~Schober, and H.~V. Poor, ``{NOMA}-based coexistence of near-field and far-field massive {MIMO} communications,'' \emph{{IEEE} Wireless Commun. Lett.}, vol.~12, no.~8, pp. 1429--1433, 2023.

\bibitem{ding2023nf2}
Z.~Ding, ``Resolution of near-field beamforming and its impact on {NOMA},'' \emph{{IEEE} Wireless Commun. Lett.}, vol.~13, no.~2, pp. 456--460, 2024.

\bibitem{zuo2023non}
J.~Zuo, X.~Mu, and Y.~Liu, ``Non-orthogonal multiple access for near-field communications,'' \emph{arXiv preprint arXiv:2304.13185}, 2023.

\bibitem{zhu2022electromagnetic}
J.~Zhu, Z.~Wan, L.~Dai, M.~Debbah, and H.~V. Poor, ``Electromagnetic information theory: Fundamentals, modeling, applications, and open problems,'' \emph{{IEEE} Wireless Commun.}, vol.~31, no.~3, pp. 156--162, 2024.

\bibitem{zhang2022near}
X.~Zhang, H.~Zhang, and Y.~C. Eldar, ``Near-field sparse channel representation and estimation in {6G} wireless communications,'' \emph{{IEEE} Trans. Commun.}, vol.~72, no.~1, pp. 450--464, 2024.

\bibitem{zhang2022bt}
Y.~Zhang, X.~Wu, and C.~You, ``Fast near-field beam training for extremely large-scale array,'' \emph{{IEEE} Wireless Commun. Lett.}, vol.~11, no.~12, pp. 2625--2629, 2022.

\bibitem{han2012game}
Z.~Han, \emph{Game theory in wireless and communication networks: theory, models, and applications}.\hskip 1em plus 0.5em minus 0.4em\relax Cambridge University Press, 2012.

\bibitem{chen2021game}
W.~Chen, S.~Zhao, R.~Zhang, and L.~Yang, ``Generalized user grouping in {NOMA} based on overlapping coalition formation game,'' \emph{{IEEE} J. Sel. Areas Commun.}, vol.~39, no.~4, pp. 969--981, 2021.

\bibitem{kaidi2019clustering}
K.~{Wang}, J.~{Cui}, Z.~{Ding}, and P.~{Fan}, ``Stackelberg game for user clustering and power allocation in millimeter wave-{NOMA} systems,'' \emph{{IEEE} Trans. Wireless Commun.}, vol.~18, no.~5, pp. 2842--2857, 2019.

\end{thebibliography}
%%%%%%%%%%%%%%%%%%%%%%%%%%%%%%%%%%%%%%%%%%%%%%%%%
%%%%%%%%%%%%%%%%%%%%%%%%%%%%%%%%%%%%%%%%%%%%%%%%%
\end{document}